\documentclass[aps,prl,showpacs,amssymb,twocolumn]{revtex4}
\usepackage{graphicx}
\usepackage{bm,amsmath}
\begin{document}
\title{Quantum logic with weakly coupled qubits}
\author{Michael R. Geller,$^{1}$ Emily J. Pritchett,$^{1}$ Andrei 
Galiautdinov,$^{1}$ and John M. Martinis$^{2}$}

\affiliation{$^1$Department of Physics and Astronomy, University of 
Georgia, Athens, Georgia 30602, USA  \\  $^2$Department of Physics, 
University of California, Santa Barbara, California 93106, USA}

\date{June 19, 2009}

\begin{abstract}
There are well-known protocols for performing CNOT quantum logic with qubits 
coupled by particular high-symmetry (Ising or Heisenberg) interactions. 
However, many architectures being considered for 
quantum computation involve qubits or qubits and resonators coupled 
by more complicated and less symmetric interactions. Here we consider 
a widely applicable model of weakly but otherwise arbitrarily coupled 
two-level systems, and use quantum gate design techniques to derive a 
simple and intuitive CNOT construction. Useful variations and extensions 
of the solution are given for common special cases.
\end{abstract}

\maketitle

Experimental realizations of gate-based quantum computation require 
the accurate implementation of universal two-qubit operations such as the
controlled-NOT (CNOT) quantum logic gate \cite{Neilsen2000}. Finding the 
best way to achieve this for a specific experimental architecture is a principal
goal of what we refer to as {\sl quantum gate design}. The CNOT problem 
can be informally stated as follows: Specify the dimension $N$  of the 
relevant Hilbert space, and a Hamiltonian
\begin{equation}
H\big( \xi_1, \xi_2, \dots , \xi_K \big)
\label{general hamiltonian with control}
\end{equation}
with some experimental control over $K$ parameters $\xi_1, \dots , \xi_K$. 
How should the control parameters be varied to generate CNOT logic
in the computational basis $\lbrace |00\rangle , |01\rangle , |10\rangle , |11\rangle
\rbrace$? For a closed system this is a control problem in the unitary group U$(N)$. 
$N$ is not necessarily equal to 4 because the Hamiltonian might include auxiliary 
non-qubit states (not in the computational basis) that help implement the 
logic. For example, an effective strategy (see, for example, Strauch 
{\it et al.}~\cite{StrauchPRL03}) 
is to use an anticrossing of the $|11\rangle$
state with a non-computational state $| {\rm aux} \rangle$ to generate a $2 \pi$ 
rotation in the two-dimensional subspace $\lbrace |11\rangle ,  
|{\rm aux} \rangle \rbrace$.
This implements the gate ${\rm CZ} \equiv {\rm diag}(1, 1, 1, -1)$
in the computational basis, out of which a CNOT can be made by pre- and 
post-application of single-qubit Hadamards. Another important example is 
Cirac and Zoller's use of vibrational modes to mediate quantum 
logic between the internal qubit states of  trapped ions \cite{CiracPRL95}.

In the familiar U(4) case, the Hamiltonian (\ref{general hamiltonian with control})
can be written in terms of Pauli matrices and their tensor products. The resulting
{\sl coupled-qubit model} usually allows control of some of the single-qubit 
operators---enough to perform arbitrary SU(2) rotations on each qubit---and possibly of the 
qubit-qubit coupling. For certain commonly occurring forms of the qubit-qubit
interaction, including the highly symmetric cases of Ising-like 
$\sigma^z_1 \sigma^z_2$ interaction \cite{Neilsen2000} 
and Heisenberg-like ${\bm \sigma}_1 \cdot {\bm \sigma}_2$ interaction 
\cite{LossPRA98}, effective protocols for implementing CNOT
gates have been established. However, many architectures being considered
for quantum computation involve qubits or qubits and resonators 
coupled by more complicated and less symmetric interactions, or would be
more accurately modeled as such. 

Here we investigate the general problem
of weakly but otherwise arbitrarily coupled qubits, and use perturbation theory
combined with other quantum gate design techniques to derive a simple and 
widely applicable CNOT pulse construction. Useful variations and 
extensions of the basic solution are given for common special cases, and
the intuitive geometric picture we employ (related to the Weyl chamber
description used by Zhang {\it et al.}~\cite{ZhangPRA03}) will be useful elsewhere 
in the design of quantum logic gates. We assume unitary evolution, which is
sensible given the generality of our result and the wide variation in
experimental coherence times. 

Zhang and Whaley \cite{ZhangPRA05a} have addressed the problem
of two-qubit gate construction using similar methods applied to a variety of 
coupled-qubit models, but focused on steering with continuous rf control
as opposed to the short pulses considered below. One of us has recently 
investigated the implementation of CNOT gates using constant rf driving 
\cite{GaliautdinovPRA07,GaliautdinovJMP07} and moderately detuned
qubits \cite{GaliautdinovPRA09}, providing constructions complementary 
to those presented here. Time-optimal and other direct quantum
control approaches are especially useful for strongly coupled and/or strongly
driven qubits, or to optimize performance in the presence of specific decohering
and/or noisy environments \cite{KhanejaPRA01,ZanardiPRA04,GrigorenkoPRL05b,SporlPRA07}, 
but early quantum logic demonstrations might best be accomplished using the
simple perturbative protocol described here.

{\it Weakly coupled qubits}. In a wide variety of physical systems being considered 
for quantum computation, the Hamiltonian for a pair of coupled qubits can be written 
(suppressing $\hbar$) as
\begin{eqnarray}
H&=&\sum_{i=1,2} \bigg( \! -\frac{\epsilon_i}{2} \, \sigma_i^z 
+ \Omega_i \cos(\epsilon_i t + \phi_i) \, \sigma^x_i \bigg) \nonumber \\
&+& \sum_{\mu,\nu = x,y,z} J_{\mu \nu} \, \sigma_1^\mu \otimes \sigma_2^\nu ,
\label{model hamiltonian}
\end{eqnarray}
with $J_{\mu \nu}$ a $3 \! \times \! 3$ real-valued tensor (possibly adjustable). 
The Hamiltonian 
(\ref{model hamiltonian}) is written in the basis of eigenstates  ($|0\rangle$ and 
$|1\rangle$) of uncoupled qubits with energy level spacings $\epsilon_i$, and the parameters $\epsilon_i$ and $\Omega_i$ (with $\Omega_i \ll \epsilon_i$) are
assumed to be experimentally controllable. More general single-qubit control is
often available but will not be needed here. Our principal assumption is that of weak coupling: The magnitude of the $J_{\mu \nu}$ are assumed to be small compared 
with the $\epsilon_i$.

Two-qubit logic gates will be implemented by combining certain entangling
operations, performed with tuned ($\epsilon_1 \! = \! \epsilon_2$) qubits, together
with single-qubit operations performed with detuned or decoupled qubits
\cite{detuningNote}. In a 
frame rotating with the tuned qubits, the Hamiltonian (\ref{model hamiltonian}) reduces
approximately to \cite{rotatiingFrameNote}
\begin{equation}
H \approx \sum_{i=1,2} \frac{\Omega_i}{2} \big( \! \cos \phi_i \, \sigma^x_i - \sin \phi_i \, \sigma^y_i  \big) + {\cal H} ,
\label{rwa hamiltonian}
\end{equation}
where
\begin{equation}
{\cal H} \equiv  J \big(\sigma_1^x \sigma_2^x + \sigma_1^y \sigma_2^y \big) 
+ J_{zz} \, \sigma_1^z \sigma_2^z + J' \big(\sigma_1^x \sigma_2^y - \sigma_1^y \sigma_2^x \big).
\label{natural hamiltonian}
\end{equation}
Here
\begin{equation}
J \equiv \frac{J_{xx} + J_{yy}}{2} \ \ \ {\rm and} \ \ \ J' \equiv \frac{J_{xy} - J_{yx}}{2}.
\end{equation}
In the computational basis,
\begin{equation}
{\cal H} = \begin{pmatrix} J_{zz}  & 0 & 0 & 0 \cr 0 & -J_{zz}  & \gamma & 0 \cr 0 & \gamma^* & -J_{zz} & 0 \cr 0 & 0 & 0 & J_{zz}  \end{pmatrix} , 
\end{equation}
where $\gamma \equiv 2(J+iJ') = |\gamma| \, e^{i \varphi}.$
To obtain (\ref{rwa hamiltonian}) we have assumed that the $J_{\mu \nu}$ and 
$\Omega$ are small compared with the qubit frequency and have neglected the
resulting rapidly oscillating terms with vanishing time-averages (the
usual rotating-wave approximation). Although 9 coupling constants are present in 
(\ref{model hamiltonian}), only 3 parameters appear in ${\cal H}$, 
making a general analysis possible. The terms in 
(\ref{natural hamiltonian}) multiplying $J$ and $J_{zz}$ are symmetric under
qubit-label exchange, whereas the $J^\prime$ term is antisymmetric and
therefore vanishes when the physical qubits in question (and their operating 
biases) are identical. Furthermore, in the common case of $J^\prime\!=\!0$ 
(which must occur when the qubits are identical but can also occur when 
they are not), ${\cal H}$ commutes with itself at different times when $J$ and 
$J_{\rm zz}$ are time dependent, leading to additional flexibility (in the form 
of ``area'' theorems) for pulse design that we will use below. We emphasize that 
${\cal H}$ is a universal Hamiltonian, applying to any pair of tuned, weakly coupled 
qubits. Coupled-qubit models with nondiagonal single-qubit drift terms can be put in the 
form (\ref{model hamiltonian}) after transformation to the uncoupled eigenstate 
basis.

{\it Cartan decomposition}. The trajectory in U(4) that 
\begin{equation}
U = T \, e^{-i \int_0^t H \, d\tau}
\label{time-ordered exponential}
\end{equation}
traces out during Schr\"odinger evolution ($T$ is the standard time-ordering
operator) can be viewed by factoring out local (single-qubit) rotations 
$u \in {\rm SU(2) \otimes SU(2)}$. A convenient way to achieve this is to use the fact 
that any element of U(4) can be written as
\begin{equation}
U = e^{i \phi} \, u_{\rm post} \, A \, u_{\rm pre},
\label{representation theorem}
\end{equation}
with
\begin{equation}
A(x,y,z) \equiv  e^{-i(x \, \sigma_1^x \sigma_2^x +  y \, \sigma_1^y \sigma_2^y
+  z \, \sigma_1^z \sigma_2^z)},
\label{entangler definition}
\end{equation}
for some local rotations $u_{\rm pre}$ and $u_{\rm post}$, real-valued coordinates
(angles) $x$, $y$, and $z$, and global phase $\phi$. This formula can be derived by
using a Cartan decomposition of the Lie algebra $su(4)$
\cite{KhanejaPRA01,KhanejaCP01,KrausPRA01}. The central 
component $A$ has the geometrical structure of a 3-torus with period $2 \pi$ and 
characterizes the nonlocal or entangling part of $U$. By performing the 
decomposition (\ref{representation theorem}) at each time $t$ and forming the 
vector ${\vec r} \equiv (x, y, z)$, we can view the evolution of the nonlocal part of 
$U$ as a trajectory ${\vec r}(t)$ through the three-dimensional space of entanglers
(\ref{entangler definition}).
A special property of (\ref{entangler definition}) is that the generators 
$\sigma^x \! \otimes \sigma^x$, $\sigma^y \! \otimes \sigma^y$, and 
$\sigma^z \! \otimes \sigma^z$ all commute (they form the abelian Cartan subalgebra). 
The minus sign introduced into the exponent of  (\ref{entangler definition}) simplifies the
analysis in the common special case of $J^\prime \! = \! 0$.

The decomposition (\ref{representation theorem}) into an entangler $A$, local
rotations $u_{\rm pre}$ and $u_{\rm post}$, and phase factor $e^{i \phi}$ is not
unique. This means that the trajectory ${\vec r}(t)$ corresponding to some
actual physical evolution is not unique. But the different options for $A$ at
each time $t$ are evidently {\sl locally equivalent} (differing by pre- and 
post-application of local rotations and a multiplicative phase factor). Furthermore, 
in the common special case of $J^\prime \! = \! 0,$ a particularly natural continuous
solution [given in (\ref{area theorems})  below] can always be chosen which has 
the simplifying property that the local rotations and phase factor are equal to the 
identity along the entire trajectory: The local rotation and global phase angles 
vanish. We note that the usefulness of the decomposition (\ref{representation theorem}) 
goes far beyond its somewhat technical role here:  (i) In architectures where 
local operations can be performed quickly and accurately (they are ``free"), the
decomposition allows one to focus directly on the remaining nonlocal part;  (ii) 
The local rotations associated with successive gates can often be combined; And
(iii), some of the experimental error incurred when implementing an entangler---the
component that doesn't change the equivalence class---can be corrected by 
modifying the $u$'s.

\begin{figure}
\includegraphics[width=8.0cm]{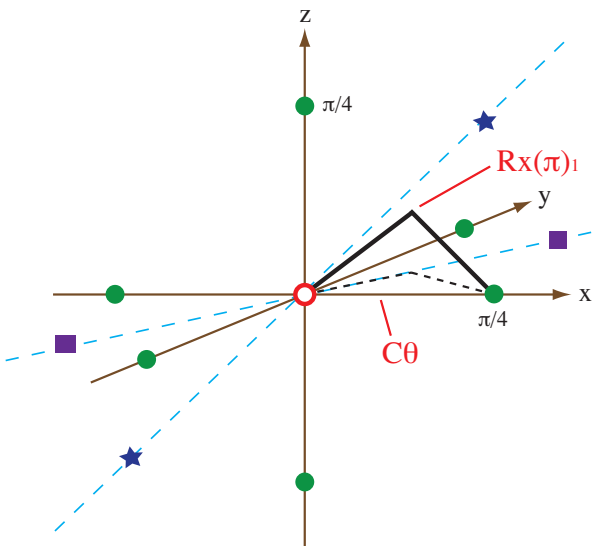} 
\caption{(color online) Three-dimensional space of entanglers 
$A$, showing the six members $A(\pm \frac{\pi}{4},0,0),$ $A(0, \pm \frac{\pi}{4},0),$ 
and $A(0,0,\pm \frac{\pi}{4})$ of the CNOT equivalence class (solid green 
circles) closest to the identity (open red circle). The Schr\"odinger evolution 
resulting from (\ref{natural hamiltonian}) with fixed positive $J$ and $J_{\rm zz}$ is 
indicated by the black trajectory, interrupted by the application of a fast $\pi$ 
pulse. The purple square at ${\vec r} = (\frac{\pi}{4}, \frac{\pi}{4},0)$
is locally equivalent to the ${\rm CNOT} \! \times \! {\rm SWAP}$ and
${\rm SWAP} \! \times \! {\rm CNOT}$ gates, and the blue star at 
${\vec r} = (\frac{\pi}{4},\frac{\pi}{4},\frac{\pi}{4})$ is locally equivalent  to the SWAP.}
\label{entangler space figure}
\end{figure} 

The concepts of local equivalence and local equivalence classes have wide
application in gate design. 
Makhlin \cite{MakhlinQIP02} has constructed an explicit formula for 3 quantities 
that can be used to test for local equivalence. The CNOT gate \cite{cnotNote}
\begin{equation}
{\rm CNOT} \equiv \begin{pmatrix} 1 & 0 & 0 & 0 \cr 0 & 1 & 0 & 0 \cr 0 & 0 & 0& 1 \cr 0 & 0 & 1 & 0 \end{pmatrix}
\label{cnot definition}
\end{equation} 
has Makhlin invariants $G_1 = 0$ and $G_2=1$ ($G_1$ is generally complex).
Two members $U$ and $U'$ of  U(4)
are locally equivalent if and only if their Makhlin invariants are identical, in which
case we write $U \sim U'.$ When restricted to a certain (nearly) tetrahedral 
region---a Weyl chamber---the angles $x,$ $y$, and $z$ are in one-to-one 
correspondence with the Makhlin invariants, leading to a unique ${\vec r}$
and a beautiful geometric description of the local equivalence classes of 
U(4) \cite{ZhangPRA03}. For our purposes, however, it will be convenient to 
work in the full toroidal space of entanglers and not restrict ${\vec r}$ to a Weyl 
chamber \cite{notationNote}.

{\it CNOT construction when $J^\prime \! = \! 0$.} 
First we consider the common special case of Hamiltonian (\ref{natural hamiltonian})
with $J^\prime \! = \! 0$, which includes the case of identical qubits. Assuming tuned 
qubits and no rf drive, the evolution (\ref{time-ordered exponential}) simplifies to $U=A$, 
with
\begin{equation}
{\vec r} = \bigg( \int J \, dt , \int J \, dt , \int J_{zz} \, dt   \bigg),
\label{area theorems}
\end{equation}
which follows a curve in the vertical plane $x=y$. The trajectory for the case of fixed 
$J$ and $J_{zz}$ is illustrated in 
Fig.~\ref{entangler space figure}. The fact that the coordinates (\ref{area theorems}) of 
the generated entangler depend only on time-integrals of the coupling constants indicates 
a type of robustness and flexibility of the associated experimental pulse sequence, analogous to the area theorem for single-qubit rotations within the rotating-wave
approximation. The closest CNOT-class entanglers are at ${\vec r} = (\pm \frac{\pi}{4},0,0)$, 
$(0,\pm\frac{\pi}{4},0)$, and $(0,0,\pm\frac{\pi}{4})$, which cannot be reached with one application of  $\, {\cal H}$ unless $J$ vanishes. In this Ising case, $A(0,0,\pm \frac{\pi}{4})$
is obtained after $\int J_{zz} \, dt = \pm \frac{\pi}{4} \, ({\rm mod} \, 2\pi)$. Generating one of these
entanglers corresponds to generating a particular member of the CNOT equivalence class;
either one is sufficient. A possible Ising pulse sequence (executed from right to left) is
${\rm CNOT} = e^{\mp i \frac{\pi}{4}} \, {\sf H}_2 \, R_z(\mp \textstyle{\frac{\pi}{2}})_1 \, 
R_z(\mp \textstyle{\frac{\pi}{2}})_2 \, A(0,0,\pm \textstyle{\frac{\pi}{4}}) \, {\sf H}_2, $
where ${\sf H} \equiv i \, R_x(\pi) \, R_y(\frac{\pi}{2})$ is a Hadamard gate \cite{rotationNote}.

Another important special case occurs when $J_{zz}$ vanishes, 
often called an $XY$ interaction \cite{steffenNote}. Here one can follow the general 
``two-shot" protocol detailed below to generate the canonical CNOT gate
(\ref{cnot definition})
or, alternatively, one can generate $A (\pm \frac{\pi}{4}, \pm \frac{\pi}{4},0)$ in a single 
shot, which is locally equivalent to both ${\rm CNOT} \! \times  \! {\rm SWAP}$ and
${\rm SWAP} \! \times \! {\rm CNOT}$, where
\begin{equation}
{\rm SWAP} \equiv \begin{pmatrix} 1 & 0 & 0 & 0 \cr 0 & 0 & 1 & 0 \cr 0 & 1 & 0& 0 \cr 0 & 0 & 0 & 1 \end{pmatrix} \! .
\label{swap definition}
\end{equation}
The gates ${\rm CNOT} \! \times \! {\rm SWAP}$ and ${\rm SWAP} \! \times \! {\rm CNOT}$
are as effective as (\ref{cnot definition}) in the sense that any quantum circuit 
written in terms of CNOTs can be immediately rewritten in terms of the swapped versions 
with no overhead \cite{cnotswapNote}. An example pulse sequence is
\begin{eqnarray}
{\rm SWAP} \! \times \! {\rm CNOT}
= \pm i \big[ R_x\big(\pm \! \textstyle{\frac{\pi}{2}}\big)_1 
\, R_x\big(\textstyle{\frac{\pi}{2}}\big)_2 \big]
\big[ R_y\big(\textstyle{\frac{\pi}{2}}\big)_1 
\nonumber \\
\times  \, R_y\big(\pm \! \textstyle{\frac{\pi}{2}}\big)_2 \big] 
 R_x\big(- \! \textstyle{\frac{\pi}{2}}\big)_2 \, 
A\big(\pm \textstyle{\frac{\pi}{4}}, \pm \textstyle{\frac{\pi}{4}},0\big)
\, R_y\big(- \! \textstyle{\frac{\pi}{2}}\big)_2  . \ \ 
\end{eqnarray}
Operations grouped together in square brackets can be performed simultaneously.
 
Although CNOT-class entanglers farther from the origin (and not shown in 
Fig.~\ref{entangler space figure}) can be reached in one shot for special 
values of $J_{zz}/J$, a faster and generally applicable protocol is to interrupt 
the evolution with a fast refocusing $\pi$ pulse applied to either qubit.
A pair of such pulses enclosing an interval of tuned qubit evolution
\begin{equation}
\cdots \, R_x(-\pi) \, e^{-i \int \! {\cal H} \, dt} R_x(\pi) \, \cdots
\end{equation}
can be viewed as transforming the interaction Hamiltonian during that interval to (note sign changes)
\begin{equation}
R_x^\dagger\big(\pi\big) \, {\cal H} \, R_x\big(\pi\big) = J \big(\sigma_1^x \sigma_2^x - \sigma_1^y \sigma_2^y \big) 
- J_{zz} \, \sigma_1^z \sigma_2^z,
\label{reflected hamiltonian}
\end{equation}
causing the reflection illustrated in Fig.~\ref{entangler space figure} and allowing the
evolution to reach any entangler on the positive $x$ axis (or negative axis for $J < 0$) . 
$R_y(\pi)$ would cause a reflection toward the $y$ axis. 

Entanglers on the $x$, $y$
and $z$ axes are locally equivalent to each other and to the controlled-phase gate
\begin{equation}
{\rm C}\theta \equiv \!
\begin{pmatrix} 1 & 0 & 0 & 0 \cr 0 & 1 & 0 & 0 \cr 0 & 0 & 1& 0 \cr 0 & 0 &  0 & e^{i\theta} \end{pmatrix}
\! \sim \! A\big(\textstyle{\frac{\theta}{4}},0,0\big) \! \sim \! 
A\big(0,\textstyle{\frac{\theta}{4}},0\big) \! \sim \!
A\big(0,0,\textstyle{\frac{\theta}{4}}\big). 
\label{controlled phase gate definition}
\end{equation} 
The entanglers at ${\vec r} = (\pm \frac{\pi}{4},0,0)$, shown in 
Fig.~\ref{entangler space figure} as solid green circles, are thus locally equivalent to 
the CZ gate, and hence to the CNOT. The identity
\begin{equation}
R_x\big(- \! \pi\big)_1 \, A\big(\pm \textstyle{\frac{\pi}{8}}, \pm \textstyle{\frac{\pi}{8}},z\big) \, R_x\big(\pi\big)_1
= A\big(\pm \textstyle{\frac{\pi}{8}}, \mp \textstyle{\frac{\pi}{8}},-z\big),
\nonumber
\end{equation} 
with $z$ arbitrary, allows us to reach 
\begin{equation}
A\big(\pm \textstyle{\frac{\pi}{8}},\mp\textstyle{\frac{\pi}{8}},-z\big) \,
A\big(\pm \textstyle{\frac{\pi}{8}},\pm \textstyle{\frac{\pi}{8}},z\big)
=A\big(\pm \textstyle{\frac{\pi}{4}},0,0\big)
\end{equation} 
after two entangling intervals, out of which a CNOT can be constructed according to
\begin{eqnarray}
{\rm CNOT} &=& e^{\mp i \frac{\pi}{4}} \, R_y\big(- \! \textstyle{\frac{\pi}{2}}\big)_1 
\big[ R_x\big(\mp \! \textstyle{\frac{\pi}{2}}\big)_1 
\, R_x\big(\mp \! \textstyle{\frac{\pi}{2}}\big)_2 \, \big]
\nonumber \\
&\times& A\big(\pm \! \textstyle{\frac{\pi}{4}},0,0\big)  \, R_y\big(\textstyle{\frac{\pi}{2}}\big)_1.
\end{eqnarray}

{\it Arbitrary $J^\prime$.} 
Here we assume Hamiltonian (\ref{natural hamiltonian}) with fixed, time-independent
values of $J$, 
$J_{zz}$, and $J^\prime$ (excluding the pure $\sigma_1^z \sigma_2^z$ Ising 
case). When $J^\prime \! \neq \! 0$ there are terms in the Hamiltonian that are 
not in the Cartan subalgebra and that break the symmetry under qubit exchange.
Such terms can be eliminated by performing a $z$ rotation on the second qubit by 
an angle $\varphi \equiv {\rm arg}(J+iJ'),$  
\begin{eqnarray}
R_z^\dagger\big(\varphi\big)_2 \, {\cal H} \, R_z\big(\varphi\big)_2 &=& \sqrt{J^2 + 
{J^\prime}^2  } \, \big(\sigma_1^x \sigma_2^x + \sigma_1^y \sigma_2^y \big) \nonumber \\
&+& J_{zz} \, \sigma_1^z \sigma_2^z,
\label{rotated natural hamiltonian}
\end{eqnarray}
allowing us to reach the CNOT-class entangler 
\begin{eqnarray}
A\big(\textstyle{\frac{\pi}{4}},0,0\big) &=& R_z\big(\!-\! \varphi\big)_2 \,
R_x\big(\!-\!\pi\big)_1\ e^{-i {\cal H} \, \Delta t} 
\nonumber \\
&\times& \, R_x\big(\pi\big)_1 \,
e^{-i {\cal H} \, \Delta t} \, R_z\big(\varphi\big)_2.
\end{eqnarray}
Here $e^{-i {\cal H} \, \Delta t}$ represents the action of bringing the
qubits into resonance for a time
\begin{equation}
\Delta t \equiv \frac{\pi}{8 \sqrt{J^2 + {J'}^2} }.
\label{entangling time}
\end{equation}
The complete pulse sequence in this case can be written as
\begin{eqnarray}
&{\rm CNOT} =  e^{i (\frac{3\pi}{4})} \big[ R_y\big(-\textstyle{\frac{\pi}{2}}\big)_1 \,
R_x\big(-\textstyle{\frac{\pi}{2}}\big)_2 \big] \, R_z\big(-\varphi \big)_2 &
 \nonumber \\
&\times \, R_x\big(\textstyle{\frac{\pi}{2}}\big)_1 \,
 e^{-i {\cal H} \, \Delta t} \, R_x\big(\pi\big)_1 \, e^{-i {\cal H} \, \Delta t}  \,
R_z\big(\varphi \big)_2 \, R_y\big(\textstyle{\frac{\pi}{2}}\big)_1. \ \ \  &
\label{CNOT from natural evolution identity}
\end{eqnarray}

In conclusion, we have shown how to implement the CNOT quantum logic gate with 
weakly but otherwise arbitrarily coupled qubits.
This work was supported by IARPA under grant W911NF-08-1-0336 and by the NSF 
under grant CMS-0404031. It is a pleasure to thank Ken Brown, Sue Coppersmith,
Andrew Sornborger, Matthias Steffen, and Robert Varley for useful discussions.

\bibliography{/Users/mgeller/Desktop/publications/bibliographies/MRGpre,/Users/mgeller/Desktop/publications/bibliographies/MRGbooks,/Users/mgeller/Desktop/publications/bibliographies/MRGgroup,/Users/mgeller/Desktop/publications/bibliographies/MRGqc-josephson,/Users/mgeller/Desktop/publications/bibliographies/MRGqc-architectures,/Users/mgeller/Desktop/publications/bibliographies/MRGqc-general,endnotes}

\begin{thebibliography}{23}
\expandafter\ifx\csname natexlab\endcsname\relax\def\natexlab#1{#1}\fi
\expandafter\ifx\csname bibnamefont\endcsname\relax
  \def\bibnamefont#1{#1}\fi
\expandafter\ifx\csname bibfnamefont\endcsname\relax
  \def\bibfnamefont#1{#1}\fi
\expandafter\ifx\csname citenamefont\endcsname\relax
  \def\citenamefont#1{#1}\fi
\expandafter\ifx\csname url\endcsname\relax
  \def\url#1{\texttt{#1}}\fi
\expandafter\ifx\csname urlprefix\endcsname\relax\def\urlprefix{URL }\fi
\providecommand{\bibinfo}[2]{#2}
\providecommand{\eprint}[2][]{\url{#2}}

\bibitem[{\citenamefont{Neilsen and Chuang}(2000)}]{Neilsen2000}
\bibinfo{author}{\bibfnamefont{M.~A.} \bibnamefont{Neilsen}} \bibnamefont{and}
  \bibinfo{author}{\bibfnamefont{I.~L.} \bibnamefont{Chuang}},
  \emph{\bibinfo{title}{Quantum Computation and Quantum Information}}
  (\bibinfo{publisher}{Cambridge University Press},
  \bibinfo{address}{Cambridge, England}, \bibinfo{year}{2000}).

\bibitem[{\citenamefont{Strauch et~al.}(2003)\citenamefont{Strauch, Johnson,
  Dragt, Lobb, Anderson, and Wellstood}}]{StrauchPRL03}
\bibinfo{author}{\bibfnamefont{F.~W.} \bibnamefont{Strauch}},
  \bibinfo{author}{\bibfnamefont{P.~R.} \bibnamefont{Johnson}},
  \bibinfo{author}{\bibfnamefont{A.~J.} \bibnamefont{Dragt}},
  \bibinfo{author}{\bibfnamefont{C.~J.} \bibnamefont{Lobb}},
  \bibinfo{author}{\bibfnamefont{J.~R.} \bibnamefont{Anderson}},
  \bibnamefont{and} \bibinfo{author}{\bibfnamefont{F.~C.}
  \bibnamefont{Wellstood}}, \bibinfo{journal}{Phys. Rev. Lett.}
  \textbf{\bibinfo{volume}{91}}, \bibinfo{pages}{167005}
  (\bibinfo{year}{2003}).

\bibitem[{\citenamefont{Cirac and Zoller}(1995)}]{CiracPRL95}
\bibinfo{author}{\bibfnamefont{J.~I.} \bibnamefont{Cirac}} \bibnamefont{and}
  \bibinfo{author}{\bibfnamefont{P.}~\bibnamefont{Zoller}},
  \bibinfo{journal}{Phys. Rev. Lett.} \textbf{\bibinfo{volume}{74}},
  \bibinfo{pages}{4091} (\bibinfo{year}{1995}).

\bibitem[{\citenamefont{Loss and DiVincenzo}(1998)}]{LossPRA98}
\bibinfo{author}{\bibfnamefont{D.}~\bibnamefont{Loss}} \bibnamefont{and}
  \bibinfo{author}{\bibfnamefont{D.~P.} \bibnamefont{DiVincenzo}},
  \bibinfo{journal}{Phys. Rev. A} \textbf{\bibinfo{volume}{57}},
  \bibinfo{pages}{120} (\bibinfo{year}{1998}).

\bibitem[{\citenamefont{Zhang et~al.}(2003)\citenamefont{Zhang, Vala, Sastry,
  and Whaley}}]{ZhangPRA03}
\bibinfo{author}{\bibfnamefont{J.}~\bibnamefont{Zhang}},
  \bibinfo{author}{\bibfnamefont{J.}~\bibnamefont{Vala}},
  \bibinfo{author}{\bibfnamefont{S.}~\bibnamefont{Sastry}}, \bibnamefont{and}
  \bibinfo{author}{\bibfnamefont{K.~B.} \bibnamefont{Whaley}},
  \bibinfo{journal}{Phys. Rev. A} \textbf{\bibinfo{volume}{67}},
  \bibinfo{pages}{042313} (\bibinfo{year}{2003}).

\bibitem[{\citenamefont{Zhang and Whaley}(2005)}]{ZhangPRA05a}
\bibinfo{author}{\bibfnamefont{J.}~\bibnamefont{Zhang}} \bibnamefont{and}
  \bibinfo{author}{\bibfnamefont{K.~B.} \bibnamefont{Whaley}},
  \bibinfo{journal}{Phys. Rev. A} \textbf{\bibinfo{volume}{71}},
  \bibinfo{pages}{052317} (\bibinfo{year}{2005}).

\bibitem[{\citenamefont{Galiautdinov}(2007{\natexlab{a}})}]{GaliautdinovPRA07}
\bibinfo{author}{\bibfnamefont{A.}~\bibnamefont{Galiautdinov}},
  \bibinfo{journal}{Phys. Rev. A} \textbf{\bibinfo{volume}{75}},
  \bibinfo{pages}{052303} (\bibinfo{year}{2007}{\natexlab{a}}).

\bibitem[{\citenamefont{Galiautdinov}(2007{\natexlab{b}})}]{GaliautdinovJMP07}
\bibinfo{author}{\bibfnamefont{A.}~\bibnamefont{Galiautdinov}},
  \bibinfo{journal}{J. Math. Phys.} \textbf{\bibinfo{volume}{48}},
  \bibinfo{pages}{112105} (\bibinfo{year}{2007}{\natexlab{b}}).

\bibitem[{\citenamefont{Galiautdinov}(2009)}]{GaliautdinovPRA09}
\bibinfo{author}{\bibfnamefont{A.}~\bibnamefont{Galiautdinov}},
  \bibinfo{journal}{Phys. Rev. A} \textbf{\bibinfo{volume}{79}},
  \bibinfo{pages}{042316} (\bibinfo{year}{2009}).

\bibitem[{\citenamefont{Khaneja et~al.}(2001)\citenamefont{Khaneja, Brockett,
  and Glaser}}]{KhanejaPRA01}
\bibinfo{author}{\bibfnamefont{N.}~\bibnamefont{Khaneja}},
  \bibinfo{author}{\bibfnamefont{R.}~\bibnamefont{Brockett}}, \bibnamefont{and}
  \bibinfo{author}{\bibfnamefont{S.~J.} \bibnamefont{Glaser}},
  \bibinfo{journal}{Phys. Rev. A} \textbf{\bibinfo{volume}{63}},
  \bibinfo{pages}{032308} (\bibinfo{year}{2001}).

\bibitem[{\citenamefont{Zanardi and Lloyd}(2004)}]{ZanardiPRA04}
\bibinfo{author}{\bibfnamefont{P.}~\bibnamefont{Zanardi}} \bibnamefont{and}
  \bibinfo{author}{\bibfnamefont{S.}~\bibnamefont{Lloyd}},
  \bibinfo{journal}{Phys. Rev. A} \textbf{\bibinfo{volume}{69}},
  \bibinfo{pages}{022313} (\bibinfo{year}{2004}).

\bibitem[{\citenamefont{Grigorenko and Khveshchenko}(2005)}]{GrigorenkoPRL05b}
\bibinfo{author}{\bibfnamefont{I.~A.} \bibnamefont{Grigorenko}}
  \bibnamefont{and} \bibinfo{author}{\bibfnamefont{D.~V.}
  \bibnamefont{Khveshchenko}}, \bibinfo{journal}{Phys. Rev. Lett.}
  \textbf{\bibinfo{volume}{95}}, \bibinfo{pages}{110501}
  (\bibinfo{year}{2005}).

\bibitem[{\citenamefont{Sp\"orl et~al.}(2007)\citenamefont{Sp\"orl,
  Schulte-Herbr\"uggen, Glaser, Bergholm, Storcz, Ferber, and
  Wilhelm}}]{SporlPRA07}
\bibinfo{author}{\bibfnamefont{A.}~\bibnamefont{Sp\"orl}},
  \bibinfo{author}{\bibfnamefont{T.}~\bibnamefont{Schulte-Herbr\"uggen}},
  \bibinfo{author}{\bibfnamefont{S.~J.} \bibnamefont{Glaser}},
  \bibinfo{author}{\bibfnamefont{V.}~\bibnamefont{Bergholm}},
  \bibinfo{author}{\bibfnamefont{M.~J.} \bibnamefont{Storcz}},
  \bibinfo{author}{\bibfnamefont{J.}~\bibnamefont{Ferber}}, \bibnamefont{and}
  \bibinfo{author}{\bibfnamefont{F.~K.} \bibnamefont{Wilhelm}},
  \bibinfo{journal}{Phys. Rev. A} \textbf{\bibinfo{volume}{75}},
  \bibinfo{pages}{012302} (\bibinfo{year}{2007}).

\bibitem[{det()}]{detuningNote}
\bibinfo{note}{When $J_{\rm zz} \neq 0$, the interaction between detuned qubits
  is {\sl not} suppressed, and it is preferable to switch off the coupling
  during single-qubit operations in this case.}

\bibitem[{rot({\natexlab{a}})}]{rotatiingFrameNote}
\bibinfo{note}{Transformation to a frame rotating with qubit frequencies
  $\epsilon_1$ and $\epsilon_2$ is defined according to $|\psi\rangle
  \rightarrow \chi \, |\psi\rangle,$ where $\chi \equiv \prod_{i}
  R_z\big(\textstyle{ \int} \epsilon_i \, dt \big)_i.$ Under this
  transformation, $ \sigma_i^\mu \rightarrow \chi \, \sigma_i^\mu \,
  \chi^\dagger = \sum_{\nu} {\cal R}_z^{\mu \nu}\big(\textstyle{ \int}
  \epsilon_i \, dt \big) \, \sigma_i^\nu, $ where \begin{equation} {\cal
  R}_z(\alpha) \equiv \begin{pmatrix} \cos \alpha & \sin \alpha & 0 \cr -\sin
  \alpha & \cos \alpha & 0 \cr 0 & 0 & 1 \end{pmatrix} \nonumber \end{equation}
  is an SO(3) rotation matrix.}

\bibitem[{\citenamefont{Khaneja and Glaser}(2001)}]{KhanejaCP01}
\bibinfo{author}{\bibfnamefont{N.}~\bibnamefont{Khaneja}} \bibnamefont{and}
  \bibinfo{author}{\bibfnamefont{S.~J.} \bibnamefont{Glaser}},
  \bibinfo{journal}{Chem. Phys.} \textbf{\bibinfo{volume}{267}},
  \bibinfo{pages}{11} (\bibinfo{year}{2001}).

\bibitem[{\citenamefont{Kraus and Cirac}(2001)}]{KrausPRA01}
\bibinfo{author}{\bibfnamefont{B.}~\bibnamefont{Kraus}} \bibnamefont{and}
  \bibinfo{author}{\bibfnamefont{J.~I.} \bibnamefont{Cirac}},
  \bibinfo{journal}{Phys. Rev. A} \textbf{\bibinfo{volume}{63}},
  \bibinfo{pages}{062309} (\bibinfo{year}{2001}).

\bibitem[{\citenamefont{Makhlin}(2002)}]{MakhlinQIP02}
\bibinfo{author}{\bibfnamefont{Y.}~\bibnamefont{Makhlin}},
  \bibinfo{journal}{Quant. Info. Proc.} \textbf{\bibinfo{volume}{1}},
  \bibinfo{pages}{243} (\bibinfo{year}{2002}).

\bibitem[{cno({\natexlab{a}})}]{cnotNote}
\bibinfo{note}{The CNOT operator defined in (\ref{cnot definition}) specifies
  qubit 1 to be the control qubit. The locally equivalent alternative
  controlling the second qubit is ${\rm SWAP} \! \times \! {\rm CNOT} \! \times
  \! {\rm SWAP}$.}

\bibitem[{not()}]{notationNote}
\bibinfo{note}{We also adopt a slightly different notation than that of
  Ref.~[\onlinecite{ZhangPRA03}], writing ${\vec c}$ as $-2{\vec r}$.}

\bibitem[{rot({\natexlab{b}})}]{rotationNote}
\bibinfo{note}{$R_\mu(\theta)_i \equiv e^{-i(\theta/2)\sigma^\mu_i}$ is a $\mu$
  rotation on qubit $i$.}

\bibitem[{ste()}]{steffenNote}
\bibinfo{note}{Pulse sequences for the $XY$ case were previously derived in
  unpublished notes by Matthias Steffen.}

\bibitem[{cno({\natexlab{b}})}]{cnotswapNote}
\bibinfo{note}{These gates are also equivalent {\sl double} CNOTs, consisting
  of a pair targeting each qubit in succession (see \cite{cnotNote}):
  \begin{eqnarray} {\rm SWAP} \! \times \! {\rm CNOT} &=& {\rm CNOT} \! \times
  \! \big( {\rm SWAP} \, {\rm CNOT} \, {\rm SWAP} \big) \nonumber \\ {\rm CNOT}
  \! \times \! {\rm SWAP} &=& \big( {\rm SWAP} \, {\rm CNOT} \, {\rm SWAP}
  \big) \! \times \! {\rm CNOT}. \nonumber \end{eqnarray}}.

\end{thebibliography}

\end{document}